\documentclass[11pt]{cernrep}
\usepackage{graphicx}
\usepackage{epsfig}
\usepackage{wrapfig}
\usepackage{cite,./mcite}

\newcommand{\pom}{{I\!\!P}}
\newcommand{\reg}{{I\!\!R}}
\newcommand{\xpom}{x_\pom}

\newcommand{\lapprox}{\stackrel{<}{_{\sim}}}
\newcommand{\gapprox}{\stackrel{>}{_{\sim}}}
\newcommand{\alphapom}{\alpha_{_{\rm I\!P}}}
\newcommand{\mx}{M_{_{\rm X}}}
\newcommand{\my}{M_{_{\rm Y}}}
\newcommand{\sigrd}{\sigma_r^{D(3)}}

\begin{document}

\title{HERA Diffractive Structure Function Data
       and Parton Distributions 
\footnote{ \ \ Contributed to the Proceedings
of the Workshop on HERA and the LHC, DESY and CERN, 2004-2005.}
}

\author{
Paul Newman$^a$,
Frank-Peter Schilling$^b$
}

\institute{
$^a$ School of Physics and Astronomy, University of Birmingham, 
B15 2TT, United Kingdom \\
$^b$ CERN/PH, CH-1211 Geneva 23, Switzerland \\
}

\maketitle

\begin{abstract}
Recent diffractive structure function
measurements by the H1 and ZEUS experiments at HERA
are reviewed. 
Various data sets, obtained  
using systematically different selection and reconstruction methods,
are compared. NLO DGLAP QCD fits are performed to the most precise 
H1 and ZEUS data and diffractive parton densities are obtained in each 
case. Differences between the $Q^2$ dependences of the H1 and 
ZEUS data are reflected as differences between the diffractive
gluon densities. 
\end{abstract}

\section{Introduction}
\label{sec:intro}

In recent years, several new measurements of the semi-inclusive
`diffractive' deep inelastic scattering (DIS)
cross section for the process $ep \rightarrow eXY$ at HERA 
have been released by the H1 and ZEUS 
experiments \cite{Chekanov:2005vv,Chekanov:2004hy,h1ichep02,h1ichep02mb,h1eps03,h1ichep02fps}. The data
are often presented in the form of a $t$-integrated reduced diffractive 
neutral current cross section
$\sigma_{r}^{D(3)}$, defined through\footnote{For a full definition of 
all terms and variables used, see for example \cite{h1ichep02}.}
\begin{equation}
\frac{d^3\sigma^{ep \rightarrow eXY}}{{\rm d} x_\pom \ {\rm d} x \ {\rm d} Q^2} = \frac{4\pi
  \alpha^2}{x Q^4}\left ( 1 - y + \frac{y^2}{2} \right )
\sigma_{r}^{D(3)}(\xpom,x,Q^2) \ ,
\label{sigmar}
\end{equation}
or in terms of a diffractive
structure function $F_2^{D(3)}(x_\pom,\beta,Q^2)$.
Neglecting any contributions from $Z^0$ exchange,
\begin{equation}
\sigma_r^{D(3)} = F_2^{D(3)} - \frac{y^2}{1+(1-y)^2} F_L^{D(3)} \ ,
\label{eq:sigf2fl}
\end{equation}
such that $\sigma_r^{D(3)} = F_2^{D(3)}$ is a good approximation except
at very large $y$. The new data span a wide kinematic
range, covering several orders of magnitude in $Q^2$, $\beta$ and $x_\pom$.  

Within the framework of QCD hard scattering collinear
factorisation in diffractive DIS \cite{Collins:1997sr},
these data provide important constraints on the diffractive parton
distribution functions (dpdf's) of the proton. These
dpdf's are a crucial input for calculations of the cross sections
for less inclusive diffractive processes in DIS, such as dijet or 
charm production \cite{H1:jets,H1:charm}. 
In contrast to the case of inclusive scattering, the dpdf's extracted
in DIS are not expected to be directly applicable to hadron-hadron 
scattering \cite{Collins:1992cv,Berera:1994xh,Berera:1995fj,Collins:1997sr}.
Indeed, diffractive factorisation breaks down spectacularly when HERA dpdf's
are applied to diffractive proton-proton interactions at the 
TEVATRON \cite{Affolder:2000vb}. It may, 
however, be possible to recover good agreement
by applying an additional `rapidity gap survival probability' 
factor to account for secondary
scattering between the beam 
remnants \cite{Bjorken:1992er,Gotsman:1998mm,Cox:1999dw,Kaidalov:2003xf}.
The HERA dpdf's thus remain
an essential ingredient in the prediction of diffractive cross sections
at the LHC, notably the diffractive Higgs cross 
section \cite{forshaw:here}.
Although the poorly known 
rapidity gap survival probability leads to the largest
uncertainty in such calculations, the 
uncertainty due to the input dpdf's also plays a significant role.
In \cite{h1ichep02}, 
the H1 collaboration made a first attempt to assess the 
uncertainty from this source, propagating the experimental errors
from the data points to the 
`H1 2002 NLO fit' parton densities and assessing the theoretical
uncertainties from various sources. 

In this contribution, we investigate the compatibility between various
different measurements of $F_2^D$ by H1 and ZEUS. 
We also apply the techniques developed
in \cite{h1ichep02} to ZEUS data in order to
explore the consequences of differences between the H1 and ZEUS 
measurements in terms of dpdf's.

\section{Diffractive Selection Methods and Data Sets Considered}

One of 
the biggest challenges in measuring diffractive cross sections, and 
often the source of large systematic uncertainties, is
the separation of diffractive events 
in which the proton remains intact
from non-diffractive events and from proton-dissociation
processes in which the proton
is excited to form a system with a large mass, $\my$. Three distinct 
methods 
have been employed by the HERA experiments, which
select diffractive events of the type $ep \rightarrow eXY$, where $Y$
is a proton or at worst a low mass proton
excitation. These methods are complimentary
in that their systematics due to the rejection of proton dissociative and
non diffractive contributions are almost independent
of one another. They are explained in detail below.
\begin{itemize}
  \item {\bf Roman Pot Spectrometer Method.} 
Protons scattered through very small angles are detected directly in
detectors housed in `Roman Pot' insertions to the beampipe
well downstream the interaction point. The proton
4-momentum at the interaction point is reconstructed from the position and
slope of the tracks in these detectors, given a knowledge of the 
beam optics in the intervening region. 
The Roman Pot devices are known as the Leading Proton
Spectrometer (LPS) in the case of ZEUS and the Forward Proton Spectrometer 
(FPS) in H1. The Roman pot method provides the cleanest
separation between elastic, proton dissociative and non-diffractive events.
However, acceptances are rather poor, such that statistical
uncertainties are large in the data sets obtained so far. 
  \item {\bf Rapidity Gap Method.} 
This method is used by H1 for diffractive structure function measurements
and by both H1 and ZEUS for the investigation of final state observables.
The outgoing proton is not observed,
but the diffractive nature of the event is inferred from the presence of a 
large gap in the rapidity distribution of the final state hadrons,
separating the $X$ system from the unobserved $Y$ system. 
The diffractive kinematics are reconstructed from the mass of the $X$
system, which is well measured in the main detector components.
The rapidity gap must span the acceptance
regions of various
forward\footnote{The forward hemisphere corresponds to that of the outgoing
proton beam, where the pseudorapidity $\eta = - \ln \tan \theta / 2$ is
positive.} detector components. For the H1 data presented here, these detectors
efficiently identify activity in the pseudorapidity range 
$3.3 < \eta \lapprox 7.5$. 
The presence of
a gap extending to such large pseudorapidities is sufficient to ensure
that $\my \lapprox 1.6 \ {\rm GeV}$. 
In light of the poor knowledge of the $\my$
spectrum at low masses, no attempt is made to correct the data for the
small remaining proton dissociation contribution, but rather the 
cross sections are quoted integrated over $\my < 1.6 \ {\rm GeV}$. 
  \item {\bf {\boldmath $\mx$} Method.} 
Again the outgoing proton
is not observed, but rather than requiring a large rapidity gap, diffractive
events are selected on the basis of the inclusive
$\ln \mx^2$ distribution. Diffractive events are responsible for a plateau
in this distribution at low $\ln \mx^2$, such that they can be
selected cleanly for the lowest $\mx$ values. At intermediate $\mx$,
non-diffractive contributions are subtracted on the basis of a two component
fit in which the
non-diffractive component rises exponentially. 
This method is used for diffractive structure function 
measurements by ZEUS. It does not discriminate between 
elastic and low $\my$ proton-dissociative contributions. Results are quoted
for $\my < 2.3 \ {\rm GeV}$. 
\end{itemize}

\noindent
Four recent data sets are considered, for which full details
of luminosities and kinematic ranges can be found in table \ref{tab:datasets}.
\begin{itemize}
\item Published data from ZEUS taken in 1998 and 1999, 
using the $\mx$ method and taking advantage
of the increased forward acceptance offered by a 
new plug calorimeter (`ZEUS-$\mx$') \cite{Chekanov:2005vv}.
\item Published ZEUS data obtained with the LPS 
using data taken in 1997 (`ZEUS-LPS') \cite{Chekanov:2004hy}.
\item Preliminary H1 data obtained using the rapidity gap method, combining 
three measurements using different data sets 
from the period 1997-2000 for
different regions in $Q^2$ (`H1-LRG') \cite{h1ichep02,h1ichep02mb,h1eps03}. 
\item Preliminary H1 data obtained using the FPS,
based on data taken in 1999 and 
2000 (`H1-FPS') \cite{h1ichep02fps}.
\end{itemize}

\begin{table}[h]
\begin{center}
{\footnotesize
\begin{tabular}{|l|l|l|l|l|l|l|l|}
\hline
Label          & Ref.                   & Reconstruction &    Lumi                  & \multicolumn{4}{c|}{Kinematic range} \\
               &                        & Method & $\mathcal{L} [{\rm pb^{-1}}]$  & $\my [{\rm GeV}]$ & $Q^2 [{\rm GeV^2}]$ & $\beta$ & $x_\pom$    \\
\hline
\hline
ZEUS-$\mx$        & \cite{Chekanov:2005vv}               & $\mx$ method  & $4.2$         & $<2.3$ & $2.7\ldots55$   & $0.003 \ldots 0.975$ & $0.0001 \ldots 0.03$\\
ZEUS-LPS       & \cite{Chekanov:2004hy}               & Roman Pot     & $12.8$           & $M_p$  & $2.4\ldots39$   & $0.007 \ldots 0.48$& $0.0005 \ldots 0.06$\\
H1-LRG & \cite{h1ichep02,h1ichep02mb,h1eps03} & Rapidity Gap  & $3.4\ldots63$ & $<1.6$ & $1.5\ldots1600$ &$0.01 \ldots 0.9$ & $0.0001 \ldots 0.05$ \\
H1-FPS & \cite{h1ichep02fps}                  & Roman Pot     & $25$          & $M_p$  & $2.6\ldots20$   &$0.01 \ldots 0.7$ & $0.002 \ldots 0.05$\\  
\hline
\end{tabular}
}
\end{center}
\caption{Overview of the data sets discussed here. The quoted kinematic ranges in $Q^2$, $\beta$
and $x_\pom$ correspond to the bin centres.}
\label{tab:datasets}
\end{table}

\section{Comparisons between Data Sets}

In this section, 
the $\xpom$ dependences of the data from the different measurements are 
compared at fixed values of $Q^2$ and $\beta$. Since the various measurements
are generally presented at different $Q^2$ and $\beta$ values, it is
necessary to transport the data to the same values. The $\beta$ and $Q^2$
values of the H1-LRG data are chosen as the reference points. The factors
applied to data points from the other measurements
are evaluated using two different
parameterisations, corresponding to the results of
QCD fits to 1994 H1 data \cite{Adloff:1997sc} and to a subset of 
the present H1-LRG data at intermediate $Q^2$ \cite{h1ichep02}
(see also
section~\ref{sec:fits}). In order to 
avoid any significant bias arising from this procedure, 
data points are only considered further here if the correction applied is
smaller than $50\%$ in total and if the correction factors obtained from
the two parameterisations are in agreement to better than $25\%$.
In practice, these criteria only lead to the rejection of data points
in the ZEUS-$\mx$ data set at $Q^2 = 55 \ {\rm GeV^2}$ 
and $\beta = 0.975$, where the 
poorly known high $\beta$ dependence of the diffractive cross section
implies a large uncertainty on the factors required to 
transport them to $\beta = 0.9$. Elsewhere, there is reasonable agreement
between the factors obtained from the two parameterisations and no
additional uncertainties are assigned as a consequence of this procedure.

\begin{figure}[htb]
  \centering
    \epsfig{file=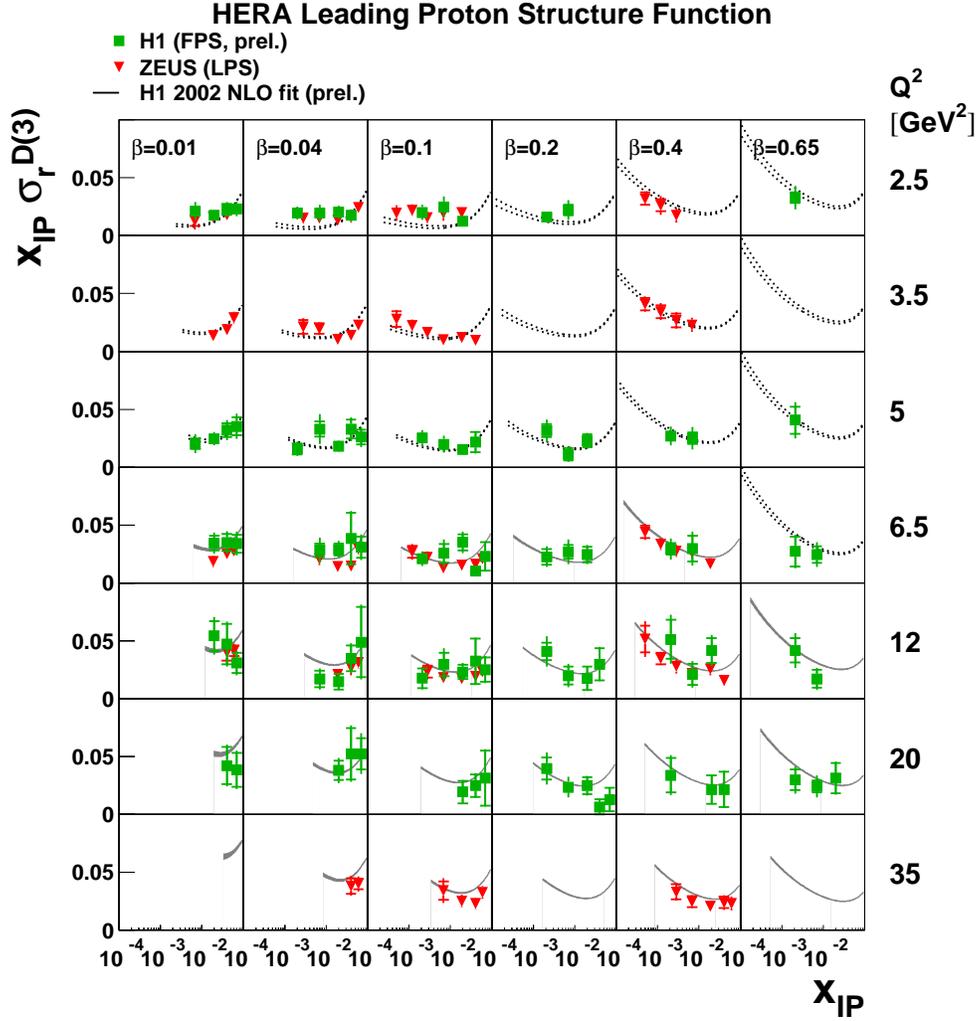,width=0.84\linewidth}
\caption{
Comparison of the Roman Pot data from H1 and ZEUS, scaled by a factor 1.1
such that they correspond to $\my < 1.6 \ {\rm GeV}$. The $Q^2$ and $\beta$
values have been shifted to the
H1-LRG bin centres using small translation factors.
The upper and lower curves form an error band on the predictions from
the H1 2002 NLO 
QCD fit to the H1-LRG data (experimental errors only). 
Dotted lines are used for kinematic 
regions which were not included in the fit.
Normalisation uncertainties of $^{+12\%}_{-10\%}$ on the ZEUS LPS data 
and $15 \%$ on the factor applied to
shift the datasets to $\my < 1.6 \ {\rm GeV}$ are
not shown. }
\label{fig:lpsfps}
\end{figure}

Since the various data sets correspond to different 
ranges in the outgoing proton system mass, $\my$, additional factors are
required before comparisons can be made. 
For all data and fit comparisons, all data are transported to
the H1 measurement range of $\my < 1.6 \ {\rm GeV}$ and 
$|t| < 1 \ {\rm GeV^2}$. 
The leading proton
data are scaled by a factor $1.1$ \cite{Adloff:1997mi} to 
correspond to the range $\my < 1.6 \rm\ GeV$ 
and the ZEUS-$\mx$ data are scaled to the same range by a 
further factor of $0.7$ \cite{Chekanov:2005vv}, 
such that the overall factor is $0.77$. 
The uncertainties on these factors are large, giving rise to 
normalisation uncertainties of perhaps $15 \%$ between the different
data sets.

\begin{figure}[htbp]
  \centering
    \epsfig{file=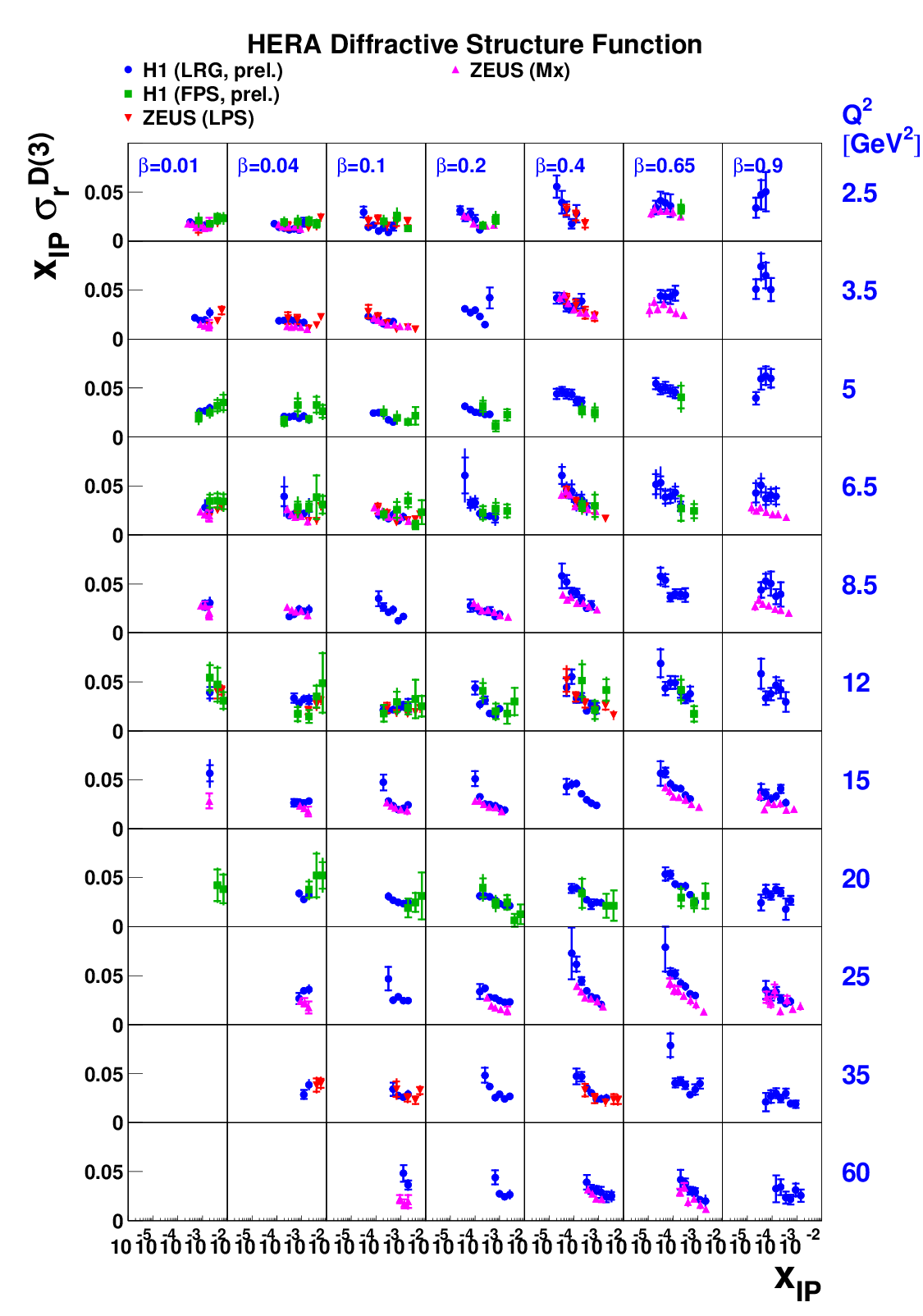,width=0.95\linewidth}
    \caption{Summary plot of all diffractive DIS data sets considered here.
Additional H1-LRG data with $Q^2 < 2.5 \ {\rm GeV^2}$,
$Q^2 = 45 \ {\rm GeV^2}$ and $Q^2 > 60 \ {\rm GeV^2}$ are not shown.
The $Q^2$ and $\beta$
values for all data sets
have been shifted to the
H1 bin centres using small translation factors. 
The ZEUS data 
have been multiplied by
a universal factor of 0.77 and the LPS and FPS data by factors
of 1.1, such that all data sets correspond to
$\my < 1.6 \ {\rm GeV}$.
Relative normalisation uncertainties of
$15\%$ due to these factors and further normalisation
uncertainties of $\pm 6.7 \%$ (H1-LRG) and 
$^{+12\%}_{-10\%}$ (ZEUS-LPS) data are not shown.}
\label{fig:allcomp}
\end{figure}

The ZEUS-LPS and H1-FPS data 
are compared in figure~\ref{fig:lpsfps}.
Within the experimental uncertainties, the two data sets are in good agreement.
Both data sets are also consistent with a parameterisation of the
H1-LRG data \cite{h1ichep02} based on the H1 
2002 NLO QCD fit, which is also shown. 
This good agreement between the H1-LRG and the
Roman Pot data is also shown explicitly in 
figure~\ref{fig:allcomp}.

\begin{figure}[htb]
  \centering
     \epsfig{file=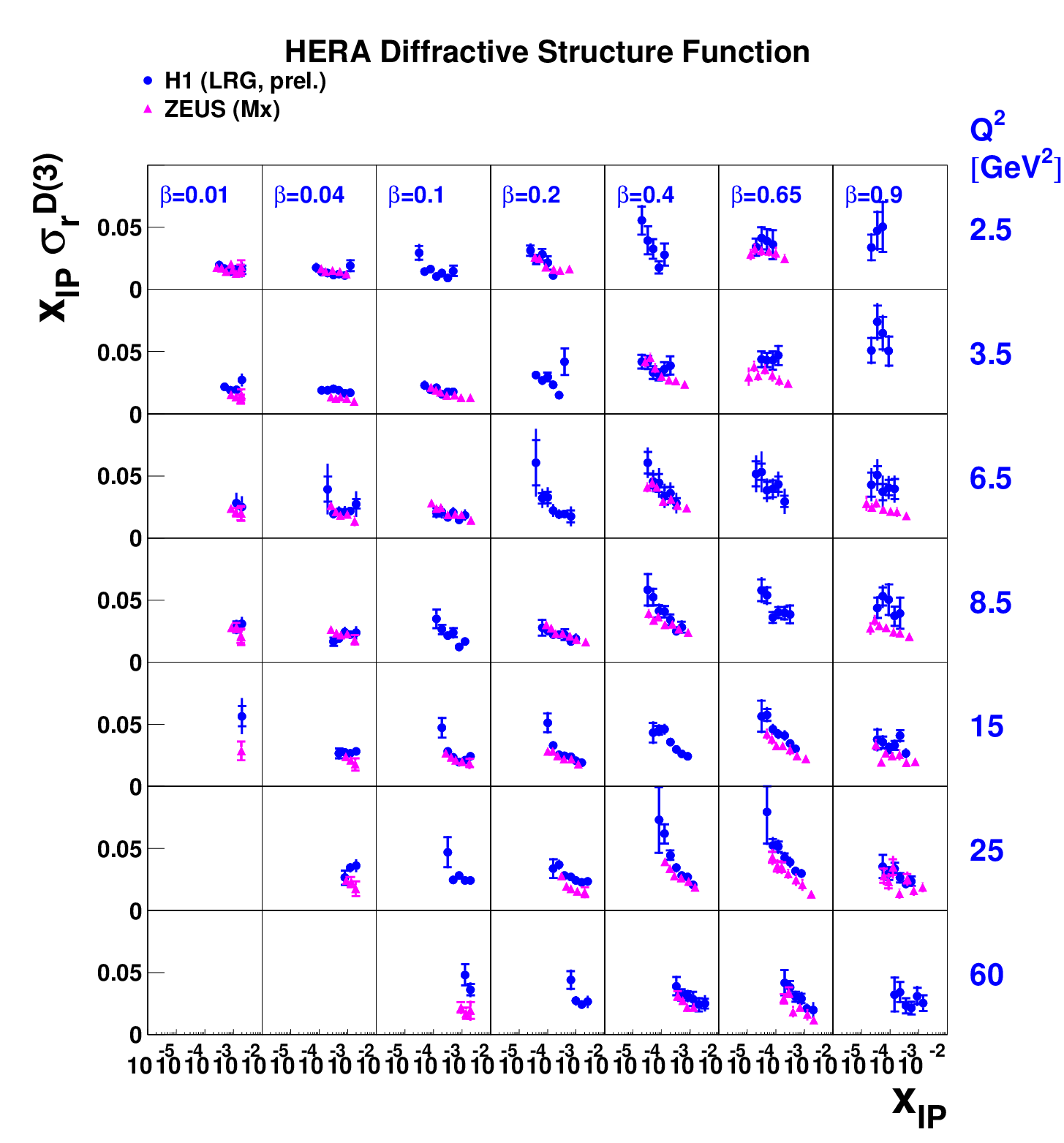,width=0.84\linewidth}
\caption{Comparison of the ZEUS-$\mx$ data with a subset of the H1-LRG 
data. The $Q^2$ and $\beta$ values of the ZEUS 
data have been shifted to the
H1 bin centres using small translation factors. The ZEUS data 
have also been multiplied by
a universal factor of 0.77, such that both data sets correspond to
$\my < 1.6 \ {\rm GeV}$. Normalisation uncertainties of
$15\%$ on this factor and of $\pm 6.7 \%$ on the H1 data are
not shown.}
\label{fig:lrgmx}
\end{figure}

In figure \ref{fig:lrgmx}, a comparison is made between the H1-LRG and 
the ZEUS-$\mx$ data after all factors have been applied. For much of the 
kinematic range, 
there is tolerable agreement between the two data sets. However,
there are clear regions of disagreement. One is at the largest $\beta$
(smallest $\mx$),
where the H1 data lie significantly above the ZEUS data for 
$Q^2 \lapprox 20 \ {\rm GeV^2}$. Another is at intermediate and low $\beta$,
where the two data sets show significantly different dependences on $Q^2$.
With the factor of 0.77 applied to the ZEUS data, there is good agreement
at low $Q^2$, but the ZEUS data lie below the H1 data at large $Q^2$. If the
factor of 0.77 is replaced with a value closer to unity, 
the agreement improves at large $Q^2$,
but the H1 data 
lie above the ZEUS data at low $Q^2$. These inconsistencies
between the different data sets are discussed further in 
section \ref{sec:fits}.

For completeness, figure \ref{fig:allcomp} shows a comparison between all four 
data sets considered. 

\section{Diffractive Parton Distributions}
\label{sec:fits}

\subsection{Theoretical Framework and Fit to H1-LRG Data}
\label{theory}

In this contribution, we adopt the fitting procedure used by
H1 in \cite{h1ichep02}, where next-to-leading order (NLO) QCD fits are
performed to diffractive reduced cross section, $\sigrd$,
data \cite{h1ichep02,H1:f2dhiq2}
with $6.5 \leq Q^2 \leq 800 \ {\rm GeV^2}$ and the $\beta$ and
$\xpom$ ranges given in table~\ref{tab:datasets}.

The proof that QCD hard scattering collinear
factorisation can be applied to diffractive DIS \cite{Collins:1997sr} implies
that in the leading $\log(Q^2)$ approximation, the cross section for
the diffractive process $e p \rightarrow e X Y$ can be written in
terms of convolutions of universal partonic cross sections
$\hat{\sigma}^{e i}$ with diffractive parton distribution functions (dpdf's)
$f_i^D$ \cite{Trentadue:1993ka,Berera:1994xh,Grazzini:1997ih}, 
representing probability distributions for a
parton $i$ in the proton under the constraint that the proton is scattered
with a particular 4 momentum.
Thus, at leading
twist,\footnote{A framework also exists to include higher order 
operators \cite{Blumlein:2002fw}.}
\begin{equation}
\frac{{\rm d^2} \sigma(x,Q^2,x_\pom,t)^{ep \rightarrow e X p^\prime}}
{{\rm d} x_\pom \ {\rm d} t} \ = \
\sum_i \int_x^{x_\pom}{\rm d}\xi \
\hat{\sigma}^{e i}(x,Q^2,\xi) \
f_i^D(\xi,Q^2,x_\pom,t) \ .
\label{equ:diffpdf}
\end{equation}
This factorisation formula is valid for sufficiently large $Q^2$ and
fixed $x_\pom$ and $t$. It also applies to the case of
proton dissociation into a system of fixed mass $\my$ and thus
to any cross section which is integrated over a fixed range in $\my$.
The partonic cross sections
$\hat{\sigma}^{e i}$ are the same as those for inclusive DIS and
the dpdf's $f_i^D$, which are not known from
first principles, should obey the 
DGLAP evolution equations \cite{Hebecker:1997gp}.

In addition to the rigorous theoretical prescription represented
by equation~(\ref{equ:diffpdf}), an additional assumption is necessary
for the H1 fits in \cite{h1ichep02}, that the shape of the dpdf's 
is independent of $\xpom$ and $t$ and that
their normalisation is controlled by Regge asymptotics \cite{Ingelman:1984ns}. 
Although this assumption has no solid basis in QCD, it is compatible with the
data fitted.
The diffractive parton distributions can then be factorised into a
term depending only on $x_\pom$ and $t$ and a term
depending only on $x$ (or $\beta$) and $Q^2$:
\begin{equation}
f_i^D(x_\pom,t,x,Q^2) = f_{\pom/p}(x_\pom,t) \cdot 
f_i^{\pom}(\beta=x/x_\pom,Q^2) \ .
\label{reggefac}
\end{equation}
Under this `Regge' factorisation assumption, 
the diffractive exchange can be treated as an
object (a `pomeron', $\pom$) with a partonic structure given by 
parton distributions $f_i^{\pom}(\beta,Q^2)$. 
The variable $\beta$ then corresponds to the fraction of 
the pomeron longitudinal
momentum carried by the
struck parton. The
`pomeron flux factor'
$f_{\pom/p}(x_\pom,t)$ 
represents the probability that a pomeron with
particular values of $x_\pom$ and $t$ couples to the proton. 

In the fit, the $\xpom$ dependence is
parameterised using a Regge flux factor 
\begin{equation}
f_{\pom/p}(\xpom, t) = A \cdot \int_{t_{cut}}^{t_{min}} 
\frac{e^{B_\pom t}}{\xpom^{2\alpha_\pom (t)-1}} \ 
{\rm d}t \ , 
\label{eq:fluxfac}
\end{equation}
where $t_{cut}=-1.0 \rm\ GeV^{2}$, $|t_{min}|$ is the minimum
kinematically allowed value of $|t|$ and the pomeron trajectory is 
assumed to be linear, 
$\alpha_\pom (t)= \alpha_\pom (0) + \alpha_\pom^\prime t$. The parameters
$B_\pom$ and $\alpha^\prime$ and their uncertainties are fixed as described
in \cite{h1ichep02}. The value of $A$ is chosen such that
the flux factor is normalised to unity at $\xpom = 0.003$.
The pomeron intercept is then obtained from the 
$\xpom$ dependence of the data and 
takes the value 
$\alphapom(0)=
1.173 \ \pm 0.018 \ \mathrm{(stat.)}  \ \pm
0.017 \ \mathrm{(syst.)}  \ ^{+ 0.063}_{-0.035} \ \mathrm{(model)}$.

The description of the
data is improved with the inclusion of an additional 
separately factorisable sub-leading
exchange with a trajectory intercept of $\alpha_\reg(0) = 0.50$
and parton densities 
taken from a parameterisation of the pion \cite{Owens:1984zj}. This
exchange contributes significantly only at low $\beta$ and large $\xpom$. 

The dpdf's are modelled in terms of a
light flavour singlet
\begin{equation}
\Sigma(z) = u(z)+d(z)+s(z)+\bar{u}(z)+\bar{d}(z)+\bar{s}(z) \ ,
\end{equation}
with $u=d=s=\bar{u}=\bar{d}=\bar{s}$ 
and a gluon distribution $g(z)$ at a starting scale 
$Q_0^2= 3 \rm\ GeV^2$.  Here, $z$ is the momentum
fraction of the parton entering the hard sub-process
with respect to the diffractive
exchange, such that $z=\beta$ 
for the lowest-order quark parton model process, whereas 
$0<\beta<z$ for higher order processes.  The
singlet quark and gluon distributions
are parameterised using the form
\begin{equation}
z p_i (z,Q_0^2) = \left [ \sum_{j=1}^n C_j^i P_j(2z-1)   \right ]^2
e^{\frac{0.01}{z-1}} \ ,
\end{equation}
where $P_j(\xi)$ is the $j^{\rm th}$ member of a set of Chebychev
polynomials\footnote{$P_1=1$, $P_2=\xi$ and $P_{j+1}(\xi)=2\xi
  P_j(\xi)-P_{j-1}(\xi)$.}. The series is squared to ensure
positivity. 
The exponential term is added to guarantee
that the dpdf's tend
to zero
in the limit of $z\rightarrow 1$. 
It has negligible influence on the extracted partons at
low to moderate $z$. 
The numbers of terms
in the polynomial parameterisations are
optimised to the precision of the data, with
the first three terms in the series used
for both the quark singlet and
the gluon distributions, yielding 3
free parameters ($C_j^\Sigma$ and $C_j^g$)  
for each.
The
normalisation of the sub-leading exchange contribution at high $\xpom$
is also determined by the fit such that the total number of free
parameters is $7$.
The data used in the fit are restricted to
$\mx > 2 \rm\ GeV$ to suppress non-leading twist contributions. 
The effects of $F_L^D$ are considered through its relation to
the NLO gluon density, such that no explicit cut on $y$ is required.

The
NLO DGLAP equations are used to evolve the dpdfs to $Q^2>Q_0^2$
using the method of
\cite{Adloff:2000qk}, extended for diffraction. 
No momentum sum rule is imposed.  Charm quarks
are treated in the massive scheme (appearing via boson gluon fusion
processes) with
$m_c=1.5\pm0.1 \rm\ GeV$. The strong coupling is set via\footnote{Although this
value is rather different from the world average, we retain it here for
consistency with previous H1 preliminary results, where it has been 
used consistently for QCD fits \cite{h1ichep02} and final state 
comparisons \cite{H1:jets,H1:charm}.} $\Lambda^{\rm
  \overline{MS}}_{\rm QCD}=200\pm30 \rm\ MeV$.
The statistical and experimental systematic errors on the data points
and their correlations
are propagated to obtain error bands for the
resulting dpdfs, which correspond to increases in the $\chi^2$ by one 
unit \cite{pascaudzomerlal}.
A theoretical error on the dpdfs
is estimated by variations of $\Lambda_{\rm QCD}$, $m_c$ and the 
parameterisation of the $x_\pom$
dependences as described in \cite{h1ichep02}. No
theoretical uncertainty is assigned for the choice of parton
parameterisation, though the results are consistent within the
quoted uncertainties if alternative approaches \cite{Martin:2002dr}
are used.
No inhomogeneous term of the type
included in \cite{watt} is considered here. The 
presence of such a term would
lead to a reduction in the gluon density extracted.

The central fit gives a good description of the data,
with a $\chi^2$ of $308.7$ for 306 degrees of
freedom.
The resulting diffractive quark singlet and gluon distributions
are shown in 
figure \ref{fig:partons}. Both extend to large fractional momenta  $z$.
Whereas the singlet distribution is well constrained by the fit,
there is a substantial uncertainty in the gluon distribution,
particularly for $z \gapprox 0.5$. 
The fraction of the exchanged momentum
carried by gluons 
integrated over the range $0.01<z<1$
is $75\pm15\%$ (total error), confirming the conclusion from earlier
work \cite{Adloff:1997sc} that diffraction is a gluon-induced phenomenon.
These dpdf's have been astonishingly 
successful in describing diffractive final state data 
in DIS such as charm \cite{H1:charm} and jet \cite{H1:jets} production,
which, being induced by boson-gluon fusion-type processes, are roughly 
proportional to the diffractive
gluon density.

\subsection{Fit to ZEUS Data}

A very similar fit to that described in section~\ref{theory} is performed
to the ZEUS-$\mx$ data and
the implications of the differences between the data 
sets to the
dpdf's are investigated.
The data are fitted in their original binning scheme, 
but are scaled to $\my < 1.6 \rm\ GeV$ using the factor of 0.77.
As for the fit to the H1 data, the first 3 terms are included in 
the polynomial expansions for the quark and gluon densities at the 
starting scale for QCD evolution. 
The same
fit program, prescription and parameters are used 
as was the case for the H1 2002 NLO fit, with the following
exceptions.
\begin{itemize}
\item ZEUS-$\mx$ data with $Q^2 > 4 \ {\rm GeV^2}$ are included in the fit,
whereas only H1 data with $Q^2>6.5\rm\ GeV^2$ are included.
It has been checked that the result for ZEUS is not altered significantly if
the minimum $Q^2$ value is increased to $6 \ {\rm GeV^2}$.
\item The quadratic sum of the statistical and systematic error is 
considered, i.e. there is no treatment of correlations between the data
points through the systematics.
\item No sub-leading Reggeon exchange component is included in the 
parameterisation. Including one does not 
improve or alter the fit significantly.
\item The Pomeron intercept is fitted together with the dpdf's, 
in contrast to the two stage process of \cite{h1ichep02}.
This does not influence the results significantly, though it does decrease
the uncertainty on $\alphapom(0)$. 
\end{itemize}
The fit describes the ZEUS-$\mx$ 
data well ($\chi^2=90$ for $131$ degrees of freedom) and yields
a value for the Pomeron intercept 
of $\alpha_\pom(0)=1.132\pm0.006$ (experimental error only). This value
is in agreement with the
H1 result if the full experimental and theoretical errors are taken into
account.
A good fit is thus obtained without any variation of 
$\alpha_\pom(0)$ with $Q^2$ or other deviation from Regge factorisation.

\begin{figure}[htb]
\centering
\epsfig{file=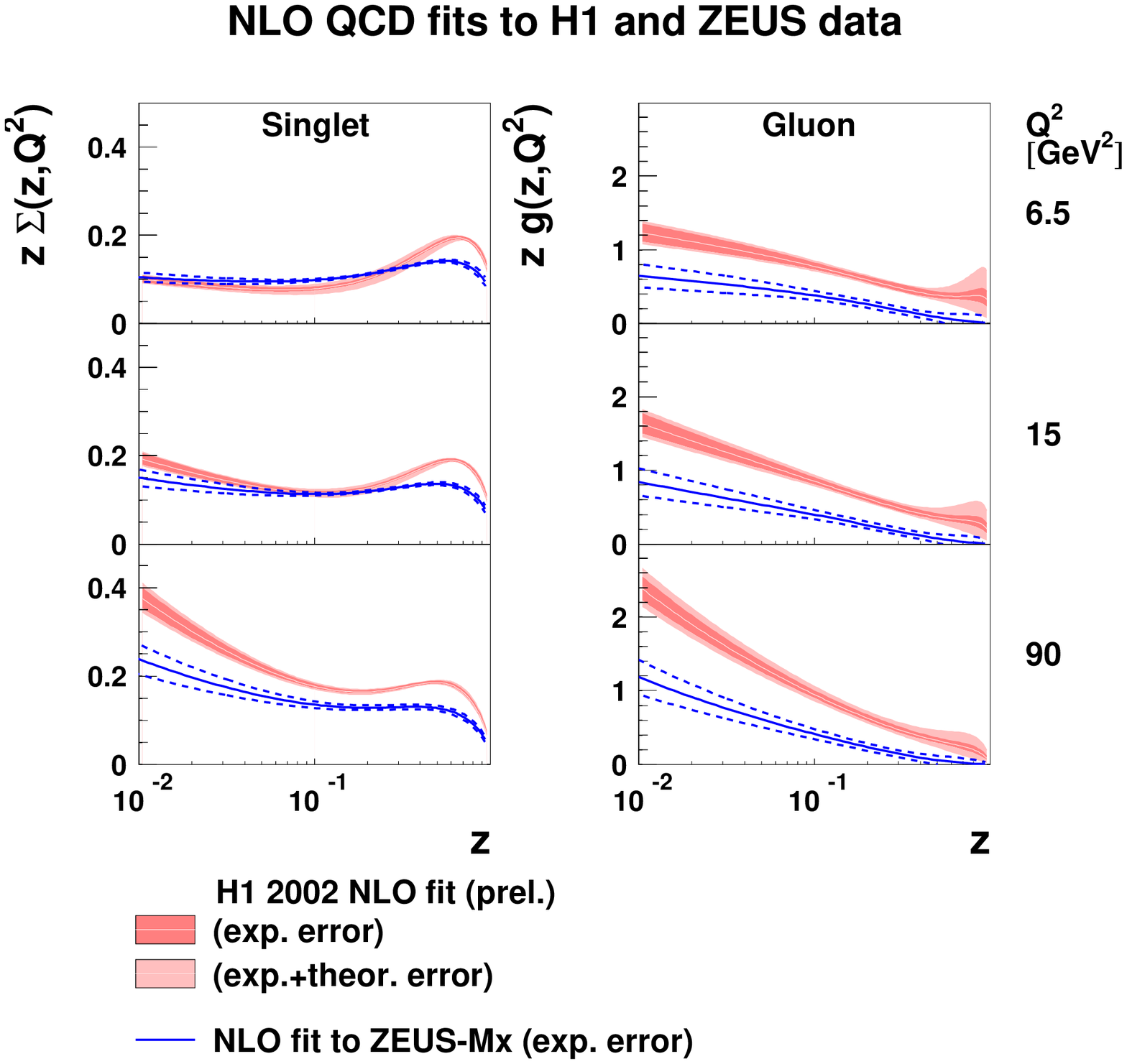,width=0.84\linewidth}
\caption{Diffractive quark singlet and 
gluon pdf's for various $Q^2$ values, as obtained 
from the NLO DGLAP fits to
the H1-LRG and ZEUS-$\mx$ data. 
The bands around the H1 result indicate the 
experimental and theoretical uncertainties. The dotted lines around the
result for ZEUS indicate the experimental uncertainty. 
The ZEUS data used in the fit are scaled by a normalisation factor
of 0.77 to match the H1-LRG range of $\my < 1.6 \ {\rm GeV}$. This factor
is reflected in the normalisations of the quark and gluon densities.
An uncertainty
of $15 \%$ on this factor is not included in the error bands shown.}
\label{fig:partons}
\end{figure}

The diffractive parton densities from the fit to the
ZEUS-$\mx$ data are compared with the results from H1 in 
figure \ref{fig:partons}. The differences observed between the H1 and 
the ZEUS data are directly reflected in the parton densities. The quark
singlet densities are closely related to the measurements of $F_2^D$
themselves. They are similar at low $Q^2$ where the H1 and ZEUS data are
in good agreement, but become different at larger $Q^2$, where 
discrepancies between the two data sets 
are observed. This difference between the 
$Q^2$ dependences of the H1 and ZEUS data is 
further reflected in a difference
of around a factor of 2 between the gluon densities, which are 
roughly proportional to
the logarithmic $Q^2$ derivative 
$\partial F_2^D / \partial \ln Q^2$ \cite{Prytz:1993vr}.

\begin{figure}[htb]
\centering
\epsfig{file=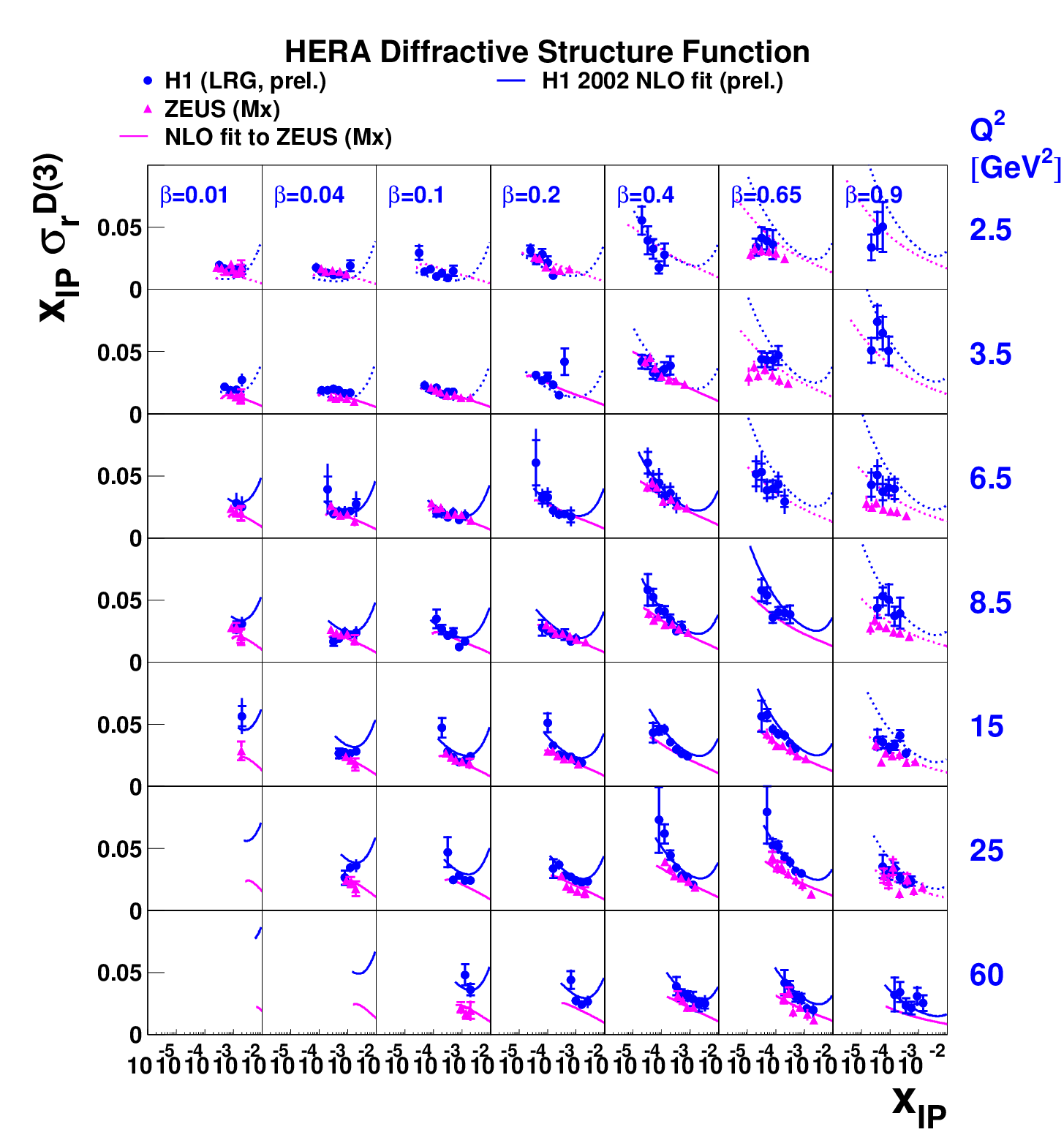,width=0.84\linewidth}
\caption{As figure \ref{fig:lrgmx}, but also showing 
the predictions using the
NLO QCD fits to the H1-LRG and ZEUS-$\mx$ data (uncertainties not shown).}
\label{fig:datafit}
\end{figure}

The H1-LRG and ZEUS-$\mx$ data are shown together with the results from
both QCD fits in figure \ref{fig:datafit}. Both fits give good descriptions
of the data from which they are obtained. The differences between the
two data sets are clearly reflected in the fit predictions, most 
notably in the $Q^2$ dependence. 

\section{Summary}

Recent diffractive structure function data from H1 and ZEUS have been 
compared directly. 
The leading proton data from both experiments (H1-FPS and ZEUS-LPS)
are in good agreement with one other and with the H1 large rapidity gap 
data (H1-LRG). There is reasonable agreement between the 
H1-LRG and the ZEUS-$\mx$ data over much of the kinematic range. However, 
differences are observed 
at the highest $\beta$ (smallest $\mx$) and
the $Q^2$ dependence at intermediate to low $\beta$
is weaker for the ZEUS-$\mx$ data than is the case for the H1-LRG data. 

An NLO DGLAP QCD fit
has been performed to the ZEUS-$\mx$ data,
using the same theoretical framework, assumptions and parameterisations
as have been employed previously for the
H1-2002-prelim NLO QCD fit to a subset of the H1-LRG data.
As a consequence of 
the differences between the $Q^2$ dependences of the H1-LRG and 
ZEUS-$\mx$ data, the 
gluon density obtained from the ZEUS data is significantly
smaller than that for H1. 

%------------------------------------------------------------------------------
%       Bibliography
%------------------------------------------------------------------------------
\bibliographystyle{heralhc} 
{\raggedright
\bibliography{proc}
}
\end{document}